\documentclass[11pt,a4paper]{article}

\usepackage[T1]{fontenc}
\usepackage[utf8]{inputenc}
\usepackage[margin=2.5cm]{geometry}
\usepackage{amsmath,amssymb}
\usepackage{bm}
\usepackage{graphicx}
\usepackage{float}
\usepackage{subcaption}
\usepackage[round,authoryear]{natbib}
\usepackage{xcolor}
\usepackage[colorlinks=true,linkcolor=blue,citecolor=blue,urlcolor=blue]{hyperref}
\usepackage{lineno}
\usepackage{booktabs}

\title{The impact of an in-situ stress gradient on the growth of a fluid-driven frictional rupture}
\author{Regina Fakhretdinova and Brice Lecampion \\
Geo-Energy Lab, EPFL
}
\date{\today}

\begin{document}
\maketitle
 
\begin{abstract}
Models of injection-induced aseismic slip on pre-existing faults commonly assume uniform in-situ stresses along the fault. They therefore predict rupture growth symmetric about the injection point. We relax this assumption and account for linear variations with depth of the resolved shear and effective normal tractions. We consider a planar fault with constant friction under plane-strain conditions, subjected to fluid injection at constant overpressure. Enforcing vanishing stress intensity factors at both rupture tips yields two coupled equations for the up-dip and down-dip tip positions. We solve them semi-analytically and compare the results with numerical simulations. The stress gradient breaks the self-similarity of the uniform-stress solution. Rupture growth is  governed by the fault-stress injection parameter $\mathcal{T}$ defined as the ratio between the initial distance to failure at the injection point and the effect of the fluid over-pressure, and a dimensionless timedefined from the ratio of the diffusion length $\sqrt{4\alpha t}$ with a stress-gradient length scale $L_c$. The latter is the ratio between the initial distance to failure at the injection point and its gradient along the fault. Growth is symmetric at early times and follows the uniform-stress solution. In the marginally pressurized regime, rupture lags behind the pressure front and the relative asymmetry grows as $1.1\,(4\alpha t/L_c^2)^{1/2}$. In the critically stressed regime, rupture outpaces fluid diffusion by a factor of order $1/\mathcal{T}$. The relevant time is then rescaled by $\mathcal{T}^2$, and the relative asymmetry grows as $\sim 0.34\,(4\alpha t/\mathcal{T}^2L_c^2)^{1/2}$. The up-dip tip reaches $\sim L_c$ earlier, and the ratio of up-dip to down-dip rupture lengths remains below two. In the marginally pressurized regime, the up-dip tip accelerates markedly compared to the down-dip tip and the relative asymmetry is stronger. 
\end{abstract}

\section{Introduction}

Fluid injection in the subsurface, notably for geothermal energy and wastewater disposal, can reactivate pre-existing faults and induce seismicity \citep{ellsworthInjectionInducedEarthquakes2013,mcgarrMaximumMagnitudeEarthquakes2014}. In-situ experiments and field observations have shown that a large part of the fault response to injection is aseismic. Induced seismicity then often appears as a by-product of an expanding aseismic slip front \citep{guglielmiSeismicityTriggeredFluid2015,cappaStabilizationFaultSlip2019,eyreRoleAseismicSlip2019,danreControlSeismicityMigration2024,danreParallelDynamicsSlow2024}. The migration of induced seismicity has long been interpreted in terms of pore-pressure diffusion \citep{shapiroEstimatingCrustPermeability1997,shapiroCharacterizationFluidTransport2002}. However, aseismic slip can outpace the pressure front, and its growth must be accounted for when interpreting seismicity migration \citep{bhattacharyaFluidinducedAseismicFault2019,danreControlSeismicityMigration2024}.

Models of fluid-driven aseismic fault slip have assumed uniform stresses along the fault plane \citep{bhattacharyaFluidinducedAseismicFault2019,viescaSelfsimilarFaultSlip2021,garagashNucleationArrestDynamic2012,saezThreedimensionalFluiddrivenStable2022a,saez2023three,saezFluiddrivenSlowSlip2024,dublanchetFluidDrivenShear2019,garagashFractureMechanicsRateandstate2021,jacqueyNucleationArrestFluidinduced2023,saezPostinjectionAseismicSlip2023}. In this paper, we extend these models to account for a systematic depth dependence of the in-situ stress field. The simplest and most analytically tractable representation is a linear variation of the shear and effective normal tractions along the fault. Before formulating this extension, it is useful to review the uniform-stress analytical models that form its basis.

A simplified 2D model provides insight into different physical regimes of slow-slip propagation \citep{viescaSelfsimilarFaultSlip2021}. Under the assumptions of a uniform initial state of stress, quasi-static linear elasticity, constant friction, and injection with constant overpressure from a line source, it was shown that the problem is governed by one dimensionless parameter which determines two limiting regimes, the so-called “marginally pressurized” and “critically stressed” regimes. Each regime corresponds to different physical mechanisms governing fault propagation (first introduced by \citet{garagashNucleationArrestDynamic2012}). The marginally pressurized limit corresponds to the case in which the fluid pressure is “just sufficient’’ to activate the fault by reducing the shear strength; a rupture propagating in this regime always lags behind the fluid front. The critically stressed limit corresponds to a fault initially close to failure; in this case, the rupture front always outpaces fluid diffusion. 

\citet{garagashNucleationArrestDynamic2012} examined the onset of dynamic rupture on a pressurized fault, incorporating linear slip-weakening friction and pore-pressure diffusion from a line source (both constant overpressure and constant injection rate cases \citep{carslawConductionHeatSolids1947}). They showed that when the initial shear traction on the fault $\tau_0$ is well below the residual shear strength $\tau_r$ at ambient conditions, rupture can be nucleated by fluid overpressure but arrests within the pressurized zone. In contrast, when the prestress is close to $\tau_r$ (i.e., the fault is critically stressed), dynamic rupture always nucleates and does not re-arrest within their formulation. 
More recently, \citet{saezThreedimensionalFluiddrivenStable2022a} extended the same problem as in \citep{garagashNucleationArrestDynamic2012,viescaSelfsimilarFaultSlip2021} to three dimensions. The study provides a comprehensive analysis of quasi-static propagation of fault slip assuming uniform in-situ stresses, constant friction, and constant fault permeability in 3D. For the constant injection-rate case (more representative of hydraulic stimulation protocols) the crack growth is self-similar and governed by a dimensionless parameter analogous to that for the constant overpressure case in 2D. Moreover, it was shown that for both critically stressed and marginally pressurized cases the rupture shape is quasi-elliptical, with aspect ratios depending on Poisson’s ratio and the dimensionless parameter.

These previous studies provide physical insight but share the simplification that the resolved tractions on the fault are taken as spatially uniform. As a result, they predict rupture shapes that are symmetric in 2D or quasi-elliptical in 3D. Field observations, however, often show asymmetric microseismic clouds, with preferential up-dip or down-dip growth depending on local stress conditions \citep{Corn16}. Such asymmetries  suggest that systematic stress variations with depth may bias the direction of rupture propagation.
In-situ stress measurements support this view. The vertical stress increases nearly linearly, typically at $23$–$27$~MPa/km in sedimentary basins and somewhat higher ($\sim$27–30~MPa/km) in crystalline basement, reflecting the weight of the overburden \citep{jaegerRockMechanicsEngineering1979,zobackReservoirGeomechanics2007}. Hydrostatic pore pressure increases at approximately $10$~MPa/km \citep{jaegerRockMechanicsEngineering1979,zobackReservoirGeomechanics2007}. Horizontal stresses are more variable, but over formation-scale depth intervals they also exhibit near-linear trends, constrained by hydraulic fracturing tests, borehole breakouts, and drilling-induced tensile fractures. Extensive datasets confirm depth dependence across tectonic settings \citep{zobackSituStressMeasurements1992,hickmanSituStressNatural1985,warpinskiInSituStressesPredominant1982,townendHowFaultingKeeps2000,heidbachWorldStressMap2019}. Although stress gradients may vary sharply across stratigraphic boundaries due to contrasts in stiffness, pore pressure, or tectonic loading, within an individual layer they are commonly well approximated by linear functions. As a result, the effective normal stress, and therefore fault strength, increases approximately linearly with depth in the absence of strong overpressure.

The observation that systematic stress gradients bias rupture growth direction is not unique to frictional fault slip. Closely related symmetry-breaking effects have long been studied in the context of tensile, fluid-driven fractures, where spatial variations in stress or pressure likewise select a preferred propagation direction. For example, \citet{weertmanTheoryWaterfilledCrevasses1971} developed an elastic theory for liquid-filled tensile cracks, originally motivated by water-filled crevasses in glaciers and later applied to vertical magma transport beneath oceanic ridges. Likewise, \citet{pollardEffectGradientsRegional1976} showed that a pressurized tensile slit subjected to linear gradients in regional stress and internal pressure acquires an asymmetric, teardrop-shaped form, with the degree of asymmetry controlled by slit length, an \emph{effective stress gradient} (difference between pressure and stress gradients), and an \emph{average driving pressure} (difference between average pressure and regional stress). Buoyancy-driven propagation of fluid-driven cracks has since been studied extensively in the context of magma transport in dykes \citep{listerFluidmechanicalModelsCrack1991,MoLe22c,MoLe23}. In that framework, a liquid-filled Mode-I crack nucleates at depth, grows in length, pinches shut at its lower tip, and subsequently rises toward the surface. The upward migration is driven by a Peach–Koehler force acting on the dislocation representation of the crack, which gives rise to a pseudo-Archimedean buoyancy force associated with gravity-induced stress and pressure gradients. In those Mode-I problems, asymmetry follows from elastic opening under spatial variations in the net normal loading on the crack faces. By contrast, the present work addresses fluid-driven \emph{frictional shear} rupture along a pre-existing interface, where pore pressure affects propagation through reduction of effective normal traction and frictional strength. While the structure of symmetry breaking by a linear gradient is analogous, the governing failure mechanism and boundary conditions differ fundamentally.

Motivated by these observations, we restrict our formulation to a single stratigraphic layer and adopt linear depth dependence for the resolved shear and effective normal tractions. This assumption preserves consistency with field data while retaining analytic and computational tractability. For a fault of fixed orientation, the depth-dependent stress tensor then maps directly into linear variations of the resolved tractions along the fault plane. In the remainder of this paper, we formulate a semi-analytical, uncoupled hydro-mechanical model with such gradients: pore-pressure diffusion is solved independently, and the resulting pressure field drives slip through quasi-static elastic equilibrium. We then derive governing conditions for rupture-tip motion and quantify how the dimensionless parameters control asymmetric slip growth.

\begin{figure}[]
\centering
\includegraphics[width=\textwidth]{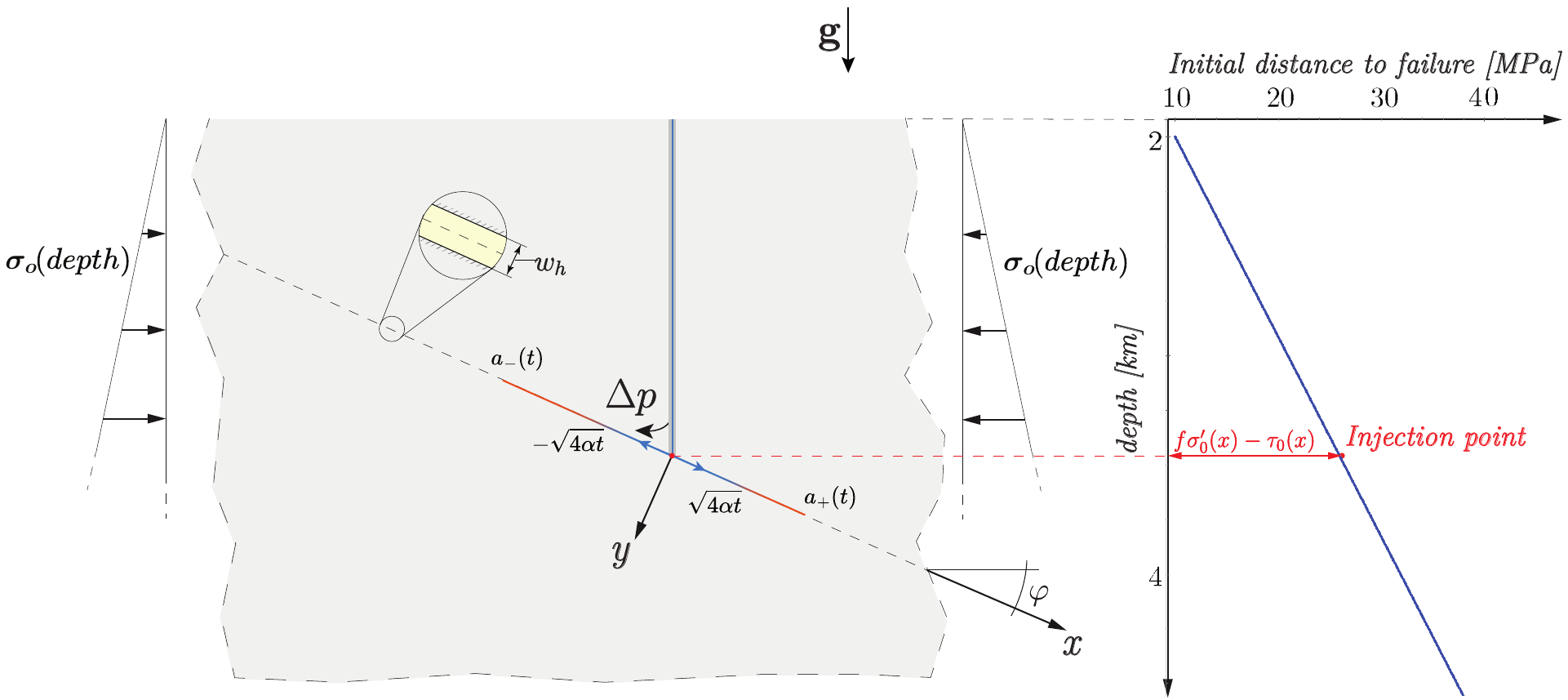}
\caption{Problem formulation in 2D. The model represents a zoom into a single stratigraphic formation layer, where the in-situ stresses can be reasonably approximated by linear gradients. The fault is treated as a zero-thickness interface embedded in an elastic, homogeneous, and isotropic solid. The slipping patch propagates asymmetrically due to the non-uniform initial tractions resolved on the fault.}
\label{fig:sktech}
\end{figure}

\section{Problem formulation}\label{sec2}

We consider a planar fault embedded in a homogeneous, isotropic, 2D linearly elastic solid (see Fig.~\ref{fig:sktech} for the in-plane case). We adopt the common fault-mechanics convention in which the effective normal traction is positive in compression.
We focus on the growth of a frictional rupture along such a planar pre-existing discontinuity driven by fluid-injection. 
We restrict to cases where the injection pressure remains below the in-situ effective normal stress, such that the fault  remains mechanically closed.
The only displacement discontinuity across the interface relates to tangential slip movement. Under 2D plane conditions, the 
displacement discontinuity across the fault reduces to a scalar tangential slip $\delta(x,t)$. 
The fault is arbitrarily oriented with respect to gravity and is characterized by a dip angle $\varphi$, a constant friction coefficient $f$, and uniform hydraulic properties (permeability $k$ and hydraulic diffusivity $\alpha$) assumed much larger than the surrounding rock. In other words, at the scale of the duration of the injection, we consider the surrounding medium impermeable. As a result, fluid flow occurs only along the fault in response to  injection from a source point at constant over-pressure $\Delta p $ above the initial pore-pressure $p_0$.

Injection of fluid inside the fault lowers the effective normal stress and reactivates the fault in shear if the over-pressure at the injection point $\Delta p$ is sufficient to reach Coulomb frictional failure expressed as a function of the shear $\tau$ and effective normal stress $\sigma^\prime$ as
\begin{equation}
    |\tau (x,t)| \le f \times \sigma^\prime(x,t).
\end{equation}

\subsection{Darcy flow along the fault}\label{sec:pressure}

We model fluid flow within the fault under the following simplifying assumptions:
\begin{itemize} 
\item the fault is a thin, permeable gouge layer of hydraulic width $w_h$ and permeability $k$, both taken constant: the fault remains closed (the effective normal traction stays compressive) and shear-induced dilatancy is neglected, so that the mechanical aperture, and thus the hydraulic properties, do not evolve with slip;
\item the permeability of the host rock is negligible compared to that of the fault over the time scales of interest, so that fluid exchange with the surrounding matrix (leak-off) is neglected;
\item the pore fluid is a single-phase, slightly compressible Newtonian fluid of viscosity $\eta$ obeying Darcy's law, and pore pressure is uniform across the fault width;
\item the initial pore pressure $p_0(x)$ is hydrostatic. Working with the overpressure $\bar p(x,t)=p(x,t)-p_0(x)$, the gravity contribution to the along-fault fluid potential cancels out.
\end{itemize}
In the plane-strain setting considered here, injection takes place through a line source perpendicular to the plane of analysis, intersecting the fault at $x=0$. Away from the source, the width-integrated fluid mass balance and Darcy's law read
\begin{equation}
w_h\,S_\epsilon\,\frac{\partial \bar p}{\partial t} + \frac{\partial w_h v}{\partial x}=0,
\qquad
v=-\frac{k}{\eta}\,\frac{\partial \bar p}{\partial x},
\label{eq:mass_darcy}
\end{equation}
where $v$ is the Darcy velocity  and $S_\epsilon =\phi_0 c_f + c_\phi$ is the storage coefficient of the fault gouge, combining the fluid compressibility $c_f$ (weighted by the initial porosity of the fault $\phi_0$) and the pore compressibility $c_\phi$. Since $w_h$, $k$ and $S_\epsilon$ are constant, \eqref{eq:mass_darcy} reduces to the linear diffusion equation
\begin{equation}
\frac{\partial \bar p}{\partial t}
= \alpha\,\frac{\partial^2 \bar p}{\partial x^2},
\qquad
\alpha = \frac{k}{S_\epsilon \,\eta},
\label{eq:diffusion}
\end{equation}
where $\alpha$ is the hydraulic diffusivity of the fault. Importantly, under these assumptions the pore-pressure field is independent of slip: the hydro-mechanical problem is one-way coupled, the pore pressure drives the mechanical response without feedback. 

We consider a constant overpressure $\Delta p$ applied at the injection point from $t=0$, with the fault initially at hydrostatic pressure and the overpressure vanishing far from the source:
\[
\bar p(x,0)=0,\qquad \bar p(0,t)=\Delta p\ (t>0),\qquad \bar p(x,t)\to 0 \ \text{as}\ |x|\to\infty .
\]
By symmetry, the problem reduces on each half-line to the classical Dirichlet problem for the 1D diffusion equation \citep{carslawConductionHeatSolids1947}, whose solution is given by:
\begin{equation}
p(x,t)-p_0(x) = \bar p(x,t) 
=
\Delta p\,
\mathrm{Erfc}\!\left(\frac{|x|}{\sqrt{4\alpha t}}\right).
\label{eq:pp_sol}
\end{equation}
The overpressure front thus propagates diffusively, with the diffusion length $\ell_d(t)=\sqrt{4\alpha t}$ as the only length scale of the fluid flow problem. The solution is self-similar. 
The corresponding injection rate (per unit length out of plane) is not constant but decays with time,
\[
Q(t) = 2\, w_h v(0^+,t)= \frac{2\,k\,w_h\,\Delta p}{\eta\sqrt{\pi\alpha t}},
\]
as expected for a constant-pressure injection. The effective normal traction on the fault then follows as $\sigma'(x,t)=\sigma'_0(x)-\Delta p\,\mathrm{Erfc}\big(|x|/\sqrt{4\alpha t}\big)$.

\subsection{Solid mechanics}
In this study, we neglect small-scale variations of the in-situ stress field with depth and focus on scales where all components of the \emph{stress tensor} vary approximately linearly with depth \citep{jaegerRockMechanicsEngineering1979}. For a fault of fixed orientation, this leads to a linear variation of the \emph{resolved} tractions along the fault plane. In the local fault coordinate system with its origin at the injection point, $Oxy$, where $x$ increases with depth (see Fig.~\ref{fig:sktech}), the initial shear $\tau_0(x)$ and effective normal $\sigma'_0(x)$ tractions on the fault can be written as 
\begin{equation}
\label{ch3:0}
\begin{aligned}
\tau_0(x)    &= \tau_0 + m_{\tau}\,x, \\
\sigma'_0(x) &= \sigma'_0 + m_{\sigma'}\,x,
\end{aligned}
\end{equation}
where $\tau_0$ and $\sigma'_0$ are the shear and effective normal tractions at the injection point, while $m_{\tau}$ and $m_{\sigma'}$ are the corresponding gradients. 

Along the fault, the local distance to failure $f\sigma_0'(x)-\tau_0(x)$ prior to injection, increases with depth for $fm_{\sigma'}-m_\tau>0$. This implies that shallower portions of the fault are closer to failure, while deeper portions are more stable. 
Such a linear  in-situ stress variation in essence introduces a limiting characteristic size. Indeed, the fault cannot be at yield initially: such a stress model is therefore valid only for $x>-L_c$, with $L_c$ the distance from the injection at which the fault would be at yield under such an in-situ stress:
\begin{equation}
    |\tau (x=-L_c,t=0)| = f \times \sigma^\prime(x=-L_c,t=0) \Rightarrow L_c =  \frac{f\times \sigma'_0-\tau_0}{f m_{\sigma'}-m_{\tau}}
\end{equation}
All further investigations are therefore intrinsically restricted to frictional rupture sufficiently smaller than $L_c$.

For an injection over-pressure $\Delta p> \sigma^\prime_0 - \tau_0/ f $ at $x=0$, a frictional rupture initiates and propagates asymmetrically due to the imposed in-situ stress gradient. Under quasi-static conditions, the shear traction $\tau(x,t)$ acting on the fault is given by the sum of the initial shear traction $\tau_0(x)$ and the change induced by the slip displacement discontinuity $\delta$ within the slipping patch/frictional crack $[-a_-(t),a_+(t)]$: 
\begin{equation}
\label{eq:stress_gr_equillibrium}
\tau(x,t)
\;=\;
\tau_0(x)
+ \frac{\mu^{\prime}}{\pi}
\int_{-a_{-}(t)}^{a_{+}(t)} 
\frac{\partial \delta(\xi,t)}{\partial \xi }\; \frac{\mathrm d\xi}{\xi-x},
\qquad -a_-(t)<x<a_+(t),
\end{equation}
where $\mu^{\prime}$ is the appropriate 2D elastic modulus (e.g.\ $\mu^{\prime}=\mu/[2(1-\nu)]$ for in-plane shear, $\mu^\prime = \mu/2 $ for anti-plane shear). Slip is defined as $\delta(x,t)=u(x,y=0^{+},t)-u(x,y=0^{-},t)$ and is taken positive in clockwise rotation.
The previous equation is nothing else than the quasi-static equilibrium for a fractured elastic solid expressed via 
the traction boundary integral equations (see \cite{hillsSolutionCrackProblems1996a,mogilevskayaLostTranslationCrack2014} for details). 
 
Inside the slipping patch  $[-a_-(t),a_+(t)]$, the Coulomb equality is satisfied: 
\begin{equation}
\tau(x,t)= f\,\sigma'(x,t),\, x \in [-a_-(t),a_+(t)]
\end{equation}
with the effective normal traction 
\begin{equation}
\sigma'(x,t) = \sigma'_0(x) - \underbrace{\big(p(x,t)-p_0(x)\big)}_{\bar p(x,t)=\Delta p \text{Erfc}(|x|/\sqrt{4 \alpha t})}.
\end{equation}
Here $p(x,t)$ is the current pore pressure, and $p_0(x)$ is the initial (hydrostatic) pore pressure, whose depth dependence is already accounted for in the background effective traction $\sigma'_0(x)$. This expression makes explicit that injection-induced \emph{overpressure} $\bar p(x,t)=p(x,t)-p_0(x)$ reduces the effective normal traction.

\subsubsection{Propagation conditions at the rupture tips}
In addition to a complete numerical simulation of the one-way coupled hydro-mechanical model, we can also obtain semi-analytically the evolution of the slipping-patch tips. A propagation condition can be obtained by recognizing that under the assumption of a constant frictional strength $f=constant$, the stress intensity factor at both rupture tips must vanish \citep{garagashNucleationArrestDynamic2012,bhattacharyaFluidinducedAseismicFault2019}. For a mode-II planar shear crack, the corresponding stress intensity factor can be obtained using the weight function approach.
\begin{equation}
\begin{aligned}
\label{eq:stress_gr_intFactors}
K_{-a_{-}(t)} &=
\frac{1}{\pi\sqrt{a_{+}(t)+a_{-}(t)}}
\int_{-a_{-}(t)}^{a_{+}(t)}
\big(\tau(\xi,t)-\tau_0(\xi)\big)\,
\frac{\sqrt{a_{+}(t)-\xi}}{\sqrt{\xi+a_{-}(t)}}\,\mathrm d\xi, \\
K_{a_{+}(t)} &=
\frac{1}{\pi\sqrt{a_{+}(t)+a_{-}(t)}}
\int_{-a_{-}(t)}^{a_{+}(t)}
\big(\tau(\xi,t)-\tau_0(\xi)\big)\,
\frac{\sqrt{\xi+a_{-}(t)}}{\sqrt{a_{+}(t)-\xi}}\,\mathrm d\xi.
\end{aligned}
\end{equation}
These expressions are obtained by inverting the integral equation \eqref{eq:stress_gr_equillibrium} for the slip density and equate it in the limits close to the rupture tip to the LEFM asymptote - see for example \cite{Rice80,viescaSelfsimilarFaultSlip2021} for details.

The right-hand side of the quasi-static elastic equilibrium \eqref{eq:stress_gr_equillibrium} relating shear stress to slip dislocation density does not prevent the shear traction at the rupture tips from becoming singular. Physically, however, the shear traction at the rupture fronts must remain finite and equal to the local shear strength. Thus, the stress intensity factors must vanish:
\begin{equation}
\label{eq:stressintensity}
K_{-a_{-}(t)}=0,
\qquad
K_{a_{+}(t)}=0.
\end{equation}
Knowing the evolution of the loads $\tau-\tau_0$ in space and time, these two conditions provide equations for the propagating tips. By summing and subtracting the relations in \eqref{eq:stressintensity}, we obtain the following two equivalent equations governing the tip positions as function of the loading in the constant friction case:   
\begin{equation}
\label{ch3:5}
\int_{-a_{-}(t)}^{a_{+}(t)}
\frac{\tau(\xi,t)-\tau_0(\xi)}{\sqrt{\xi+a_{-}(t)}\sqrt{a_{+}(t)-\xi}}
\,\mathrm d\xi
= 0,
\end{equation}
\begin{equation}
\label{ch3:6}
\int_{-a_{-}(t)}^{a_{+}(t)}
\xi\times 
\frac{\tau(\xi,t)-\tau_0(\xi)}{\sqrt{\xi+a_{-}(t)}\sqrt{a_{+}(t)-\xi}}
\,\mathrm d\xi
= 0.
\end{equation}
One can note that the first equation encapsulates the zero-order moment of the load, whereas the second equation encapsulates its first-order moment (via the term $\xi \times (\tau-\tau_0)$). For symmetric loading, the rupture is therefore necessarily symmetric: the second equation is trivially satisfied. 

In addition, in the slipping patch, Coulomb failure is satisfied, such that:
\begin{equation}
 \tau(x,t)-\tau_0(x) = f \sigma'_0(x)-\tau_0(x) -f \bar p (x,t) \qquad x\in[-a_{-},a_{+}]
 \label{eq:shear_C_inside}
\end{equation}
with the over-pressure $\bar p (x,t) $ solution of the flow problem.  

For the case of a linear stress gradient \eqref{ch3:0}, upon  substituting the
Coulomb failure condition \eqref{eq:shear_C_inside}, we easily integrate analytically the effect of the initial stress, such that the propagation conditions become:
\begin{multline}
\label{ch3:7}
\pi\big(f\sigma_{0}'-\tau_{0}\big)
+\frac{\pi}{2}\big(a_{+}(t)-a_{-}(t)\big)\big(f m_{\sigma'}-m_{\tau}\big)
\\
= f\times 
\int_{-a_{-}(t)}^{a_{+}(t)}
\frac{ \bar p(\xi,t)}
{\sqrt{\xi+a_{-}(t)}\sqrt{a_{+}(t)-\xi}}
\,\mathrm d\xi
,
\end{multline}
\begin{multline}
\label{ch3:8}
\frac{\pi}{2}\big(a_{+}(t)-a_{-}(t)\big)\big(f\sigma'_{0}-\tau_{0}\big)
+\frac{\pi}{8}\big(3a_{+}^{2}(t)-2a_{-}(t)a_{+}(t)+3a_{-}^{2}(t)\big)\big(fm_{\sigma'}-m_{\tau}\big)
\\
= f\times 
\int_{-a_{-}(t)}^{a_{+}(t)}
\xi\,
\frac{\bar p(\xi,t) }
{\sqrt{\xi+a_{-}(t)}\sqrt{a_{+}(t)-\xi}}
\,\mathrm d\xi
,
\end{multline}
where the effect of i) the initial distance to failure at the injection point $(f\sigma'_0-\tau_0)$, ii) its gradient $(f m_{\sigma'}
- m_{\tau})$ and iii) the pore-pressure increase $\bar p = p-p_0$ are properly highlighted.

In the absence of in-situ stress gradient, if the over-pressure is symmetric with respect to the injection point, the rupture is symmetric, $a_{-}=a_{+}$, and the second propagation equation is automatically satisfied.

\section{Scaling and regimes of growth}\label{sec:scaling}

\subsection{Symmetric ruptures when $f m_{\sigma'}-m_\tau =0$}

In the absence of an in-situ stress gradient, the solution of the one-way coupled hydromechanical problem is known analytically for the rupture evolution $a(t)$ and slip profile \citep{bhattacharyaFluidinducedAseismicFault2019,viescaSelfsimilarFaultSlip2021}.
We briefly recall here their main results. 
In the absence of stress gradient, the solution is symmetric: $a_{-}=a_+$ and self-similar: the rupture size and slip evolves in proportion to the pore-fluid diffusion lengthscale $\ell_d=\sqrt{4 \alpha t}$. The  proportionality factor $\lambda$  between the rupture size and the diffusion lengthscale depends on the ratio $\mathcal{T}$ between the initial distance to failure $f \sigma'_0-\tau_0$ and the amount of over-pressure driven the Coulomb rupture $f \Delta p$ (essentially the ratio between the factor of the left and right hand-side of \eqref{ch3:7}):
\[
    a(t) = \lambda({\mathcal{T}}) \times \underbrace{\sqrt{4 \alpha t}}_{\ell_d(t)} \qquad \mathcal{T} = (f \sigma'_0-\tau_0)/ (f \Delta p)
\]
We refer to $\mathcal{T}$ as the fault-stress injection parameter which here can take values between 0 and 1. 
After scaling the rupture size with $\ell_d$, Equation \eqref{ch3:7} can be integrated analytically to provide a relation between $\lambda$ and $\mathcal{T}$. Two end regimes emerge. For small values of $\mathcal{T}$ ($\ll 1$), the fault is said to be critically stressed and $\lambda \gg 1$, whereas for $\mathcal{T}\sim 1$, the injection overpressure is just sufficient to activate slip and $\lambda \ll 1$ (the fault is said to be marginally pressurized). 

The critically stressed (cs) regime is valid for $\mathcal{T}\lesssim 0.3$. In that regime, the rupture tip is way ahead of the pore-pressure diffusion front, and the following asymptote holds:
\[
    \lambda_{cs} \approx \frac{2}{\pi^{3/2} \mathcal{T}},
\]
the characteristic slip (at the center) scales as:
\[
\delta_{cs}(0) \approx \frac{\sqrt{4\alpha t} f \Delta p}{\mu^\prime} \frac{2}{\pi^{3/2}} [\ln(2 \lambda) +\gamma/2],
\]
and the corresponding characteristic elastic stress associated with the slipping patch is $\tau_{el}=\mu' \delta / a= f \sigma^\prime_0 - \tau_0$.

The marginally pressurized (mp) regime is valid for  $\mathcal{T}\gtrsim 0.7$, and in that regime:
    \[
    \lambda_{mp} \approx \frac{\pi^{3/2}}{4}(1-\mathcal{T}),
    \]
 the characteristic slip (at the center) scales as:
\[
\delta_{mp}(0) \approx  \frac{2}{\pi^{3/2}} \lambda^2 \frac{ \sqrt{4\alpha t} f \Delta p}{\mu^\prime},
\]
and the corresponding characteristic elastic stress is $\tau_{el}=\mu' \delta / a= \lambda f \Delta p$, much smaller than $f \Delta p$. In this regime, the rupture is primarily driven by the direct pore-pressure effect on fault strength with minimal impact of elastic stress transfer.

\subsection{Asymmetric ruptures when $f m_{\sigma'}-m_\tau \ne 0$}

A spatially linearly varying initial stress field breaks strict self-similarity in relation to the intrinsic length-scale $L_c= (f\sigma'_0-\tau_0)/ ( f m_{\sigma' } - m_\tau) $ associated with the in-situ stress gradient. In other words, the up- and down-dip tip positions $a_{-}$ and $a_{+}$ do not strictly follow the pore-pressure diffusion front $\ell_d(t)$. 
Nevertheless, we can still define the up and down-dip amplification factors $\lambda_{\pm}$ which now, also depends on time:
\begin{equation}
    a_{\pm} (t) = \lambda_{\pm} (\mathcal{T},t) \times \sqrt{4 \alpha t} 
\end{equation}
where here the fault-stress injection parameter $\mathcal{T}$ is defined as previously but strictly speaking from the initial distance to failure at the injection point $x=0$.

Introducing this scaling into the propagation conditions, we obtain the two coupled equations relating $\lambda_{-},\,\lambda_+$ to $\mathcal{T}$ and  a dimensionless stress gradient number $\mathcal{S}(t)$:
\begin{eqnarray}
    \pi \mathcal{T}
+\frac{\pi}{2}\big(\lambda_{+} -\lambda_{-} \big) \underbrace{\frac{\sqrt{4\alpha t} (f m_{\sigma'}-m_\tau)}{f \Delta p}}_{\mathcal{S}(t)}
= 
\int_{-\lambda_{-} }^{\lambda_{+} } \frac{ \text{Erfc}( |s|) }{\sqrt{s+\lambda_{-} }\sqrt{\lambda_{+} -s}}
\,\mathrm d s \label{eq:dim_propa_1}
\\
 \frac{\pi}{2}\big(\lambda_{+} -\lambda_{-} \big) \mathcal{T}
+\frac{\pi}{8}\big(3 \lambda_{+}^{2} -2 \lambda_{-} \lambda_{+} +3\lambda_{-}^{2} \big)  \underbrace{\frac{\sqrt{4\alpha t} (f m_{\sigma'}-m_\tau)}{f \Delta p}}_{\mathcal{S}(t)} \nonumber
\\
=  
\int_{-\lambda_{-} }^{\lambda_{+} }
s \times
\frac{\text{Erfc}(|s|) }
{\sqrt{s+\lambda_{-} }\sqrt{\lambda_{+} -s}}
\,\mathrm ds,
\label{eq:dim_propa}
\end{eqnarray}
with 
\[
\mathcal{T} = \frac{f \sigma'_0-\tau_0}{f \Delta p} \qquad
\mathcal{S}(t) = \frac{ f m_{\sigma'}-m_\tau}{f \Delta p} \times \sqrt{4 \alpha t}=
\mathcal{T} \frac{\sqrt{4\alpha t}}{L_c}
\]
and 
\[
L_c =(f \sigma'_0-\tau_0 )/(f m_{\sigma'}-m_\tau).\] The amplification factors  $\lambda_{-},\,\lambda_+$ are thus function of both $\mathcal{T}$ and $\mathcal{S}(t)$.

We observe that the in-situ stress gradient effect is captured by the ratio between the diffusion length-scale and the stress gradient scale (second terms of the left-hand sides). In obtaining \eqref{eq:dim_propa}, we have scaled the equations using the characteristic stress scale $f \Delta p$ - valid in the marginally pressurized regime (we could have alternatively scaled with $f \sigma'_0-\tau_0$).

\paragraph{Marginally pressurized regime $\mathcal{T} \sim 1 $}
In that regime, the rupture size lies well within the pore-pressure disturbed zone $\lambda \ll 1$. 
As $\mathcal{S}$ increases, asymmetry kicks in. 
Because $\mathcal{T} \sim 1 $, the impact of the stress gradient relates directly to $\sqrt{4\alpha t}/L_c$, and we can rewrite its evolution via a dimensionless time:
\[
    \mathcal{S}(t)= \mathcal{T} \sqrt{\bar{t}}\qquad \bar{t}= \frac{ 4 \alpha t }{L_c^2}
\]
Because in marginally pressurized conditions $\mathcal{T}\sim 1$, we have $\mathcal{S}(t)\approx \sqrt{\bar{t}} $. The over-pressure front will arrive at distance $L_c$ when $\mathcal S\sim 1$, and one has to expect the activation of another slipping patch up dip at $x=-L_c$ due to the increase of pore-pressure (and to additional stress transfer from the first slipping patch). We are not primarily interested in that second slipping patch here, and therefore focus on the asymmetry of the primary rupture patch prior to the activation of the second one.

From the dimensionless propagation conditions \eqref{eq:dim_propa}, we can anticipate a linear evolution of the asymmetry $\Delta \lambda$ of the rupture patch with $\mathcal{S}$, such that:
\[
\Delta \lambda = \lambda_{-} - \lambda_{+} \approx \beta(\mathcal{T}) \sqrt{\bar{t}}
 \]
at least at early times (when the rupture size remains significantly smaller than $L_c$).

\paragraph{Critically stressed regime $\mathcal{T}\ll 1$} \label{sec:cs_scale}
The dimensionless propagation conditions \eqref{eq:dim_propa} were obtained by using the marginally pressurized stress scale $f \Delta p$ which is not adequate for the critically stressed case.  
In the critically stressed regime the rupture grows significantly faster than the pore-pressure diffusion front, as  $\mathcal{T}\ll 1$ implies $\lambda \gg 1$, and the characteristic elastic stress scale is $f \sigma'_0-\tau_0$. 
Because the amplification factor is large $\lambda_{cs}\propto 1/\mathcal{T}$, the stress gradient effect is more appropriately captured via the ratio of the mean rupture scale over the stress gradient scale $L_c$
\[
    \langle a \rangle/L_c = \lambda_{cs} \frac{\sqrt{4\alpha t}}{L_c} \approx  \frac{\sqrt{4\alpha t}}{\mathcal{T} \times L_c}.
\]
We therefore see that a different dimensionless time $\hat{t}=\bar{t}/\mathcal{T}^2$ is more appropriate in the critically stressed regime. The two clocks are related through the stress-gradient number: $\mathcal{S}=\mathcal{T}\sqrt{\bar t}$ and $\sqrt{\hat t}=\mathcal{S}/\mathcal{T}^2$. 
Here, again as the up-dip rupture patch $a_{-}$  approaches the critical scale $L_c$, a secondary rupture will activate near $x=-L_c$, this time due to elastic stress transfer, as the pore-pressure front remains well within the central part of the primary rupture patch.

\section{Solutions} \label{sec:numerical}

We solve the problem using two distinct and complementary methods. 

\subsection{Rupture tips evolution via the coupled propagation conditions}
On one hand, we solve the two coupled non-linear equations \eqref{eq:dim_propa_1}-\eqref{eq:dim_propa} giving the up and down-dip amplification factors as function of dimensionless time $\bar{t}= 4 \alpha t / L_c^2$ for a given value of the fault-stress injection parameter $\mathcal{T}$. We discretize numerically the integrals and solve the residuals of the two equations for a given dimensionless time (and $\mathcal{T}$ value) using a root-finding scheme with the constraints $\lambda_{-}\ge \lambda_{+}$. We perform this over a range of dimensionless times, for different values of $\mathcal{T}$ from 0.01 to 0.98. 

It is worth noting that as time / $\mathcal{S}$ increases, the asymmetry increases to a point where ultimately no solutions are obtained (in relation to the rupture approaching the critical lengthscale $L_c$). 
Alternatively, it is also possible to directly tabulate $\mathcal{T},\mathcal{S}$ as function of 
$\lambda_{-},\, \lambda_{+}$ - by simply inverting the linear system \eqref{eq:dim_propa_1}-\eqref{eq:dim_propa} for  $\mathcal{T},\mathcal{S}$ at fixed $\lambda_{-},\, \lambda_{+}$. If doing so, we observe that above a given value of  $\lambda_{-},\, \lambda_{+}$, $\mathcal{S}$ decreases - reflecting an instability (time cannot decrease while the amplification factors increase).
This is another expression of the fact that due to a linear stress gradient, the fault is initially at yield for $-x>L_c$ - such that a complete unzipping up-dip ultimately occurs. 

\subsection{Numerical model}

\begin{table}[h!]
\centering 
\caption{Parameters of the numerical simulations. For simplicity, we just vary the injection over-pressure to reproduce critically stressed and marginally pressurized regimes. The $\mathcal{T}=0.5$ simulation is performed with $\Delta p=13.3$ MPa.}
\label{tab:num_params}
\begin{tabular}{lcc}
\toprule
Parameter & Critically stressed & Marginally pressurized \\
\midrule
$\mathcal{T}$ & 0.1 & 0.9 \\
Shear modulus $\mu^\prime$ & 13.02 GPa & 13.02 GPa \\
Friction coefficient $f$ & 0.6 & 0.6 \\
Effective normal traction at injection point $\sigma'_0$ & 90 MPa & 90 MPa \\
Shear traction at injection point $\tau_0$ & 50 MPa & 50 MPa \\
Gradients $m_{\sigma'}$, $m_\tau$ & 50 kPa/m, 20 kPa/m & 50 kPa/m, 20 kPa/m  \\
Injection overpressure $\Delta p$ & 66 MPa & 7.4 MPa \\
Hydraulic diffusivity $\alpha$ & $10^{-1}$ m$^2$/s &$10^{-1}$ m$^2$/s  \\
Domain $[-L_c,L_c]$ / number of elements $N$ & 800 m / 10 000 &  800 m / 10 000 \\
\bottomrule
\end{tabular}
\end{table}

In parallel to the predictions for the evolving rupture tip obtained from the propagation conditions presented above,
we perform coupled hydro-mechanical simulation of the complete fluid-driven rupture. 
We use the fully coupled hydro-mechanical solver for fluid-driven frictional and tensile ruptures on pre-existing discontinuities described in detail in \citet{LeBr27}. The solver handles general two and three-dimensional fractures in a linear elastic rock while the deformation of the fractures is modeled via a non-associated poroelastoplastic constitutive law.
Here, this solver is used in 2D plane strain. 
To match the previously discussed assumptions, we restrict here to the case of a Mohr-Coulomb yield criterion with  constant friction (no weakening), zero cohesion and zero dilation. In addition the interface elastic stiffnesses are taken sufficiently large (100 times the rock stiffness) such that the interface behaves essentially as rigid-plastic. As a result of the absence of shear-induced dilation, the fault permeability and storativity remain constant.  The numerical simulations therefore solve the same physical problem as the semi-analytical model, but do not assume \emph{a priori} a single slipping patch nor impose explicit conditions at the rupture tips: the extent of the slipping region emerges from the local enforcement of the Mohr-Coulomb criterion along the discretized fault.

In this numerical solver \citep{LeBr27}, the quasi-static  elastic equilibrium is discretized using a collocation
boundary element method for the  traction hyper-singular boundary integral equation relating the 
 displacement discontinuity along the fractures to the applied tractions \citep{crouchBoundaryElementMethods1983,hillsSolutionCrackProblems1996a}. The fault is discretized over a finite segment of length $L$ with $N$ straight elements with piece-wise constant displacement discontinuity over each element, and the elastic equilibrium is enforced at element mid-points using the plane-strain whole-space kernel. 
To account for the poro-elastoplastic constitutive behavior of the fault, 
the displacement discontinuity is split into an elastic and a plastic part. The elastic part is related to changes of effective traction through normal and shear interface stiffnesses $k_n$ and $k_s$, while plastic slip develops whenever the Mohr-Coulomb yield function $F=\tau-f\sigma'$ reaches zero (effective normal traction positive in compression). The constitutive update is performed at each collocation point with an implicit elastic-predictor/plastic-corrector (return-mapping) scheme \citep{SiHu98}.  

The width-integrated fluid mass balance \eqref{eq:mass_darcy} is discretized via a finite element method, with pressure unknowns at the element vertices \citep{LeBr27}. The mechanical and flow problems are solved with a fully implicit Newton-Raphson scheme for the increments of displacement discontinuities and pressure, using a preconditioned Krylov solver for the Jacobian system \citep{LeBr27}. The solver has been verified against semi-analytical solutions for fluid-driven frictional ruptures under uniform in-situ stress, in particular the two-dimensional solution of \citet{viescaSelfsimilarFaultSlip2021} and the three-dimensional solutions of \citet{saezThreedimensionalFluiddrivenStable2022a} \citep[see][]{LeBr27}. 

Here, the constant overpressure $\Delta p$ is imposed at the vertex located at the injection point. 
The initial effective tractions \eqref{ch3:0} are prescribed at the collocation points.  The rupture-tip positions $a_\pm(t)$ are extracted at each time step as the outermost positions of the elements at yield within the central slipping patch. The simulation is stopped when the element near the upper boundary $x=-L_c$ starts to yield. 

The parameters of the simulations for a critically stressed ($\mathcal{T}=0.1$) and marginally pressurized ($\mathcal{T}=0.9$) cases discussed further are listed in Table~\ref{tab:num_params}.

\begin{figure}
  \centering
  \includegraphics[width=0.9\textwidth]{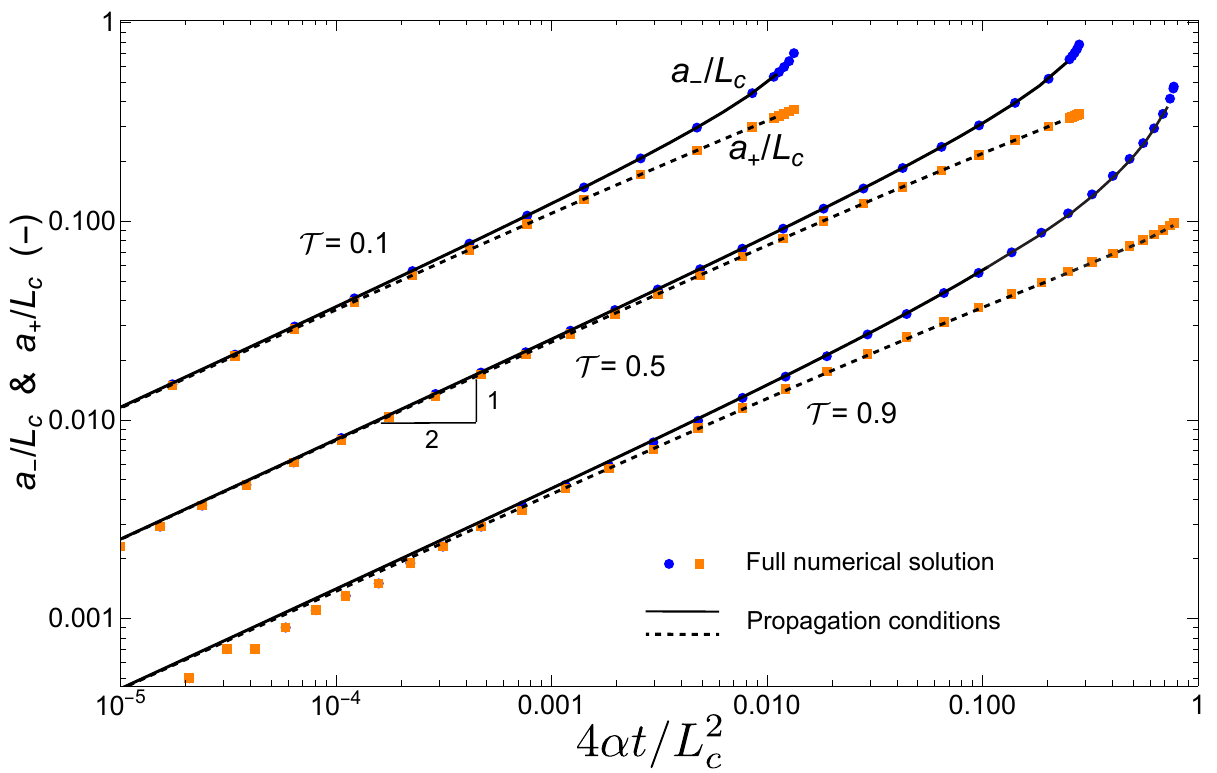}
  \caption{
  Time evolution of the up-dip ($a_{-}$) and down-dip ($a_+$) rupture tips for three different values of the fault-stress injection parameter $\mathcal{T}=0.1,\,0.5,\, 0.9$. The coupled hydro-mechanical numerical results are displayed as dots, the corresponding solution of the coupled LEFM propagation conditions are displayed as lines.
    }
    \label{fig:am_ap_t}
\end{figure}

\section{Results and discussion}\label{sec:results}

\begin{figure}
  \centering
  \includegraphics[width=0.9\textwidth]{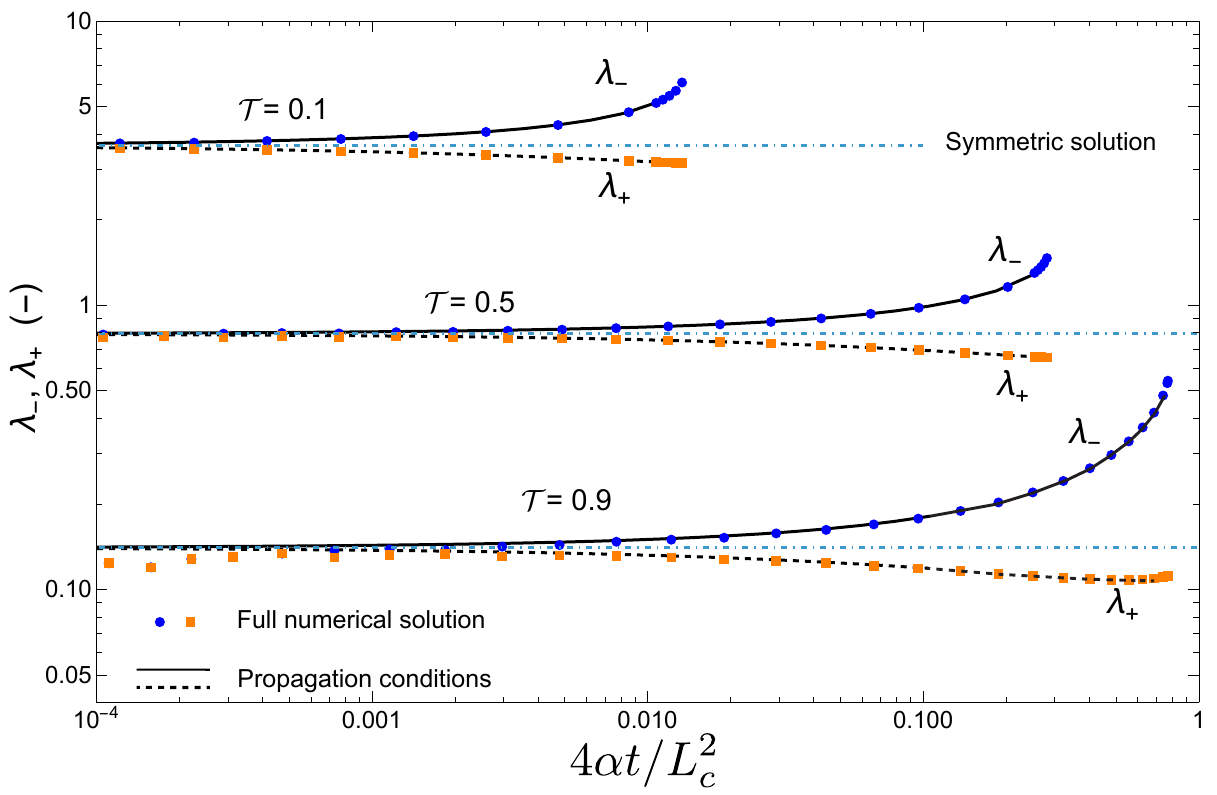}
  \caption{
  Time evolution of the up-dip ($\lambda_{-}$) and down-dip ($\lambda_+$) rupture tips amplification factor for three different values of the fault-stress injection parameter $\mathcal{T}=0.1,\,0.5,\, 0.9$. The coupled hydro-mechanical numerical results are displayed as dots, the corresponding solution of the coupled LEFM propagation conditions are displayed as  lines. The corresponding symmetric rupture values are displayed as dot-dashed blue lines.
    }
    \label{fig:lambdam_lampdap_t}
\end{figure}

We report in figure \ref{fig:am_ap_t} the evolution of the up and down dip tip positions as function of dimensionless time $\bar t$ for three different values of $\mathcal{T}$ (0.1, 0.5 and 0.9). 
Both the results of the full hydromechanical simulation (dots) and the solution of the tip propagation conditions (lines) are reported. 
The asymmetry of the rupture patch develops with time (as $\mathcal{S}$ increases) and initially follows the solution for symmetric rupture (this is better visualized on figure \ref{fig:lambdam_lampdap_t} which displays the time evolution of the amplification factors).
    The asymmetry obtained either via the solution of the propagation conditions or the complete numerical solution of the hydromechanical problem matches very well. The  numerical solution at early time is less accurate in relation to the lower number of elements at yield / in the slipping patch (the element size used is 0.08 meters).
    
   At large time, as the up dip rupture patches approach $L_c$, a secondary slipping patch activates at the up-dip end of the computational domain ($x=-L_c$) and we stop the numerical simulation. The solution of the propagation conditions (valid only for a single rupture patch) also fails to converge at similar times/$\mathcal{S}$ value.
    For all values of $\mathcal{T}$, this occurs when $a_{-}/L_c$ is of order one: between 0.5 (for $\mathcal{T}=0.9$ / marginally pressurized case) and 0.8 for $\mathcal{T}=0.5,\,0.1$ (note the log-log scale on figure \ref{fig:am_ap_t}). 
    Because the rupture grows significantly faster in the critically stressed regimes, this occurs much earlier for smaller values of $\mathcal{T}$. 
    As a result of this earlier nucleation of the secondary rupture patch (near $x=-L_c$) in the critically stressed cases, the asymmetry of the rupture appears less significant (in relative terms) than in the marginally pressurized regimes.  This can be better visualized on figure \ref{fig:lambdam_lampdap_t} where the amplification factors $\lambda_{\pm}=a_{\pm}/\sqrt{4 \alpha t}$ are reported for the same three values of $\mathcal{T}$.

\subsection{Critically stressed regime ($\mathcal{T} \ll 1$)} 

\begin{figure}
  \centering
  \includegraphics[width=\textwidth]{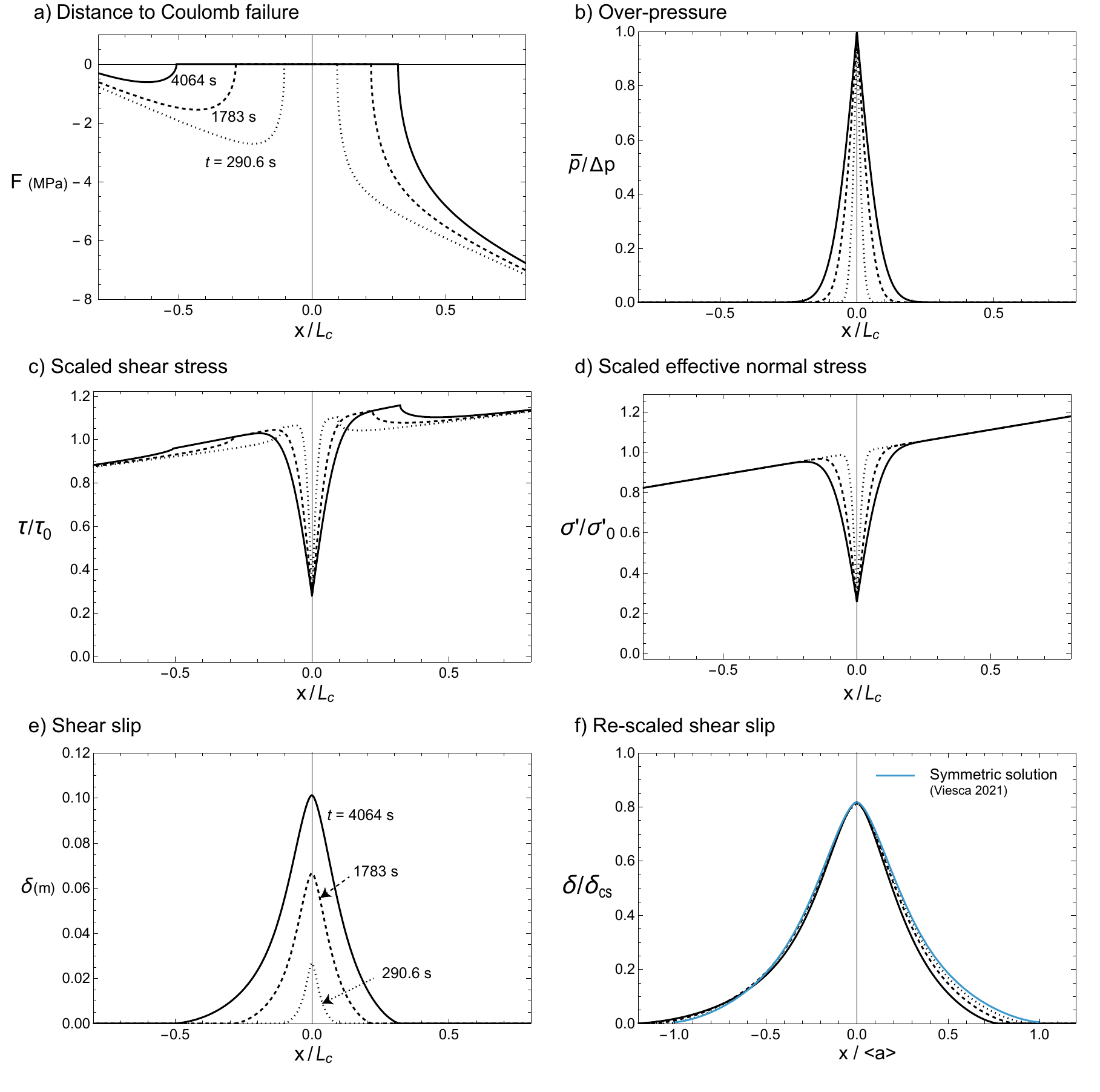}
  \caption{
  Critically stressed case ($\mathcal{T}=0.1$). Spatial profiles (at different times) along the fault scaled by $L_c$ of a) the distance to Coulomb failure $F=\tau -f \sigma^\prime$, b) the  over-pressure $\bar p =p-p_0$ scaled by the injection value, c) the shear stress scaled by its initial value at the injection point, d) the scaled effective normal stress, e) dimensional and f) rescaled shear slip (using the critically stressed scale $\delta_{cs}= \ell_d(t) f \Delta p/\mu^\prime$).
  }
  \label{fig:profiles_cs}
\end{figure}

    To better grasp the evolution of the rupture in the critically stressed regime, we display the spatial distribution along the fault of different quantities of interest at three distinct times on  figure \ref{fig:profiles_cs}: a) the distance to Coulomb failure defined as the yield function $F=\tau-f \sigma^\prime$, b) the fluid over-pressure $\bar p$ scaled by the injection value $\Delta p$, c) the shear-stress and d) effective normal stress (both scaled by their in-situ value at the injection point), e) the shear slip distribution and f) its scaled version.

    The asymmetry of the rupture is  visible on the spatial distribution of the yield function. The effect of the initial stress gradient on the distance to failure is also clear, and we can observe on figure \ref{fig:profiles_cs}a, the necessary nucleation of a secondary rupture patch (near $x=-L_c$) triggered by the growth of the primary rupture. In the critically stressed regime, the pore-pressure front lies well behind the rupture front: elastic stress transfer associated with slip is amplifying the effect of fluid-injection. 
    As already mentioned, although clearly apparent, the asymmetry of the rupture  remains small in relative terms. This is clearly visible on the slip distribution, especially when the spatial coordinate is rescaled by the mean rupture length $\langle a\rangle=(a_{-}+a_{+})/2$ and the slip scaled by the critically stressed slip scale $\delta_{cs}=\ell_d f \Delta p/\mu^\prime$. In figure \ref{fig:profiles_cs}f, the slip distribution for the symmetric case \citep{viescaSelfsimilarFaultSlip2021} is also displayed. The effect of the in-situ stress gradient remains small on this slip profile.

The fact that the asymmetry remains rather small is directly related to the  intrinsic lengthscale $L_c$ associated with the in-situ stress gradient. In the critically stressed regime, the rupture, driven primarily by elastic stress transfer, grows fast and starts to be influenced by the stress gradient at distance on par with $L_c$ - i.e. when the secondary rupture patch starts to initiate.

    \subsection{Marginally pressurized regime ($\mathcal{T} \rightarrow 1$)} 

\begin{figure}
  \centering
  \includegraphics[width=\textwidth]{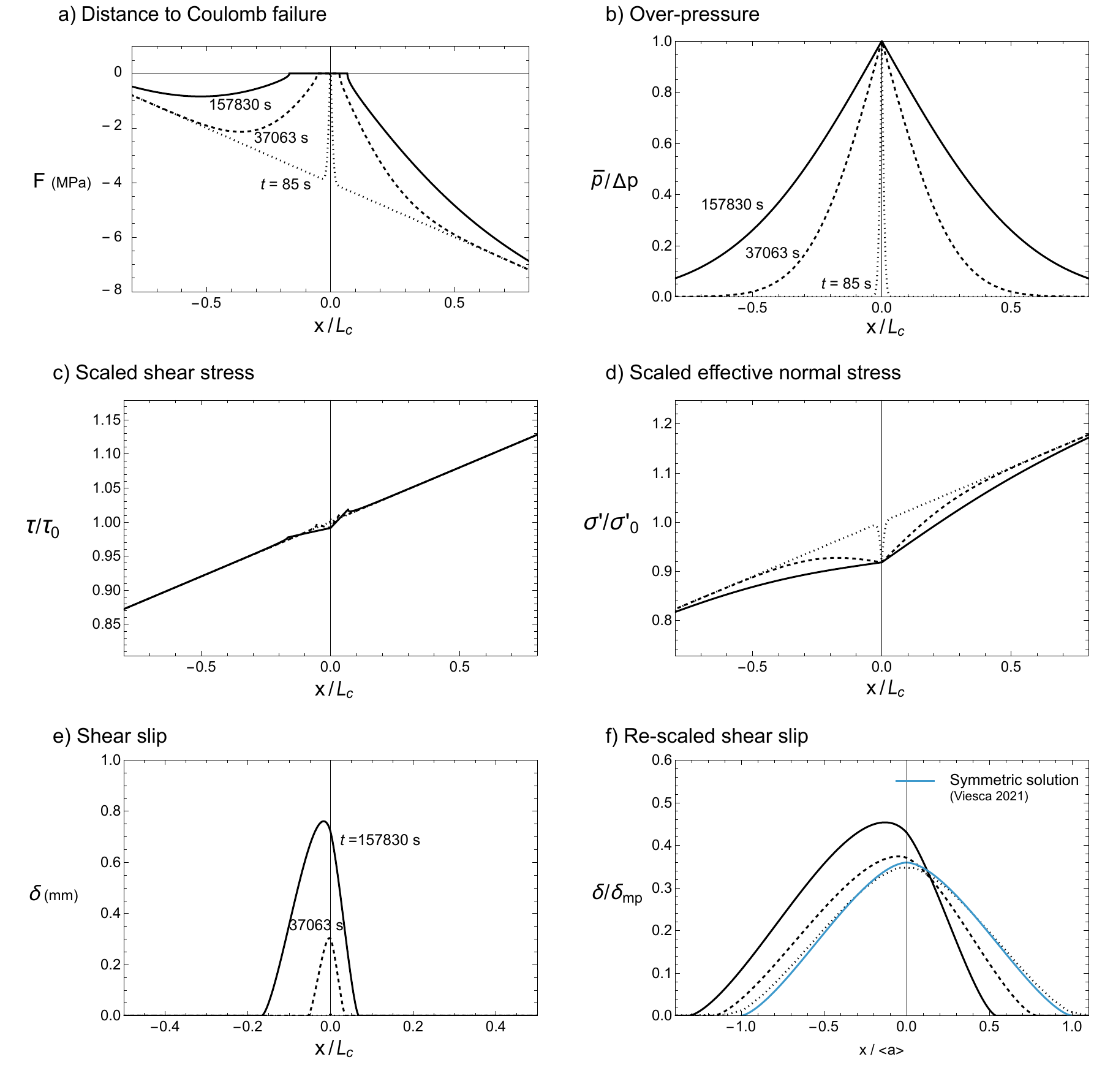}
  \caption{
  Marginally pressurized case ($\mathcal{T}=0.9$). Spatial profiles (at different times) along the fault scaled by $L_c$ of a) the distance to Coulomb failure $F=\tau -f \sigma^\prime$, b) the  over-pressure $\bar p =p-p_0$ scaled by the injection value, c) the shear stress scaled by its initial value at the injection point, d) the scaled effective normal stress, e) dimensional and f) rescaled shear slip (using the marginally pressurized scale for symmetric rupture $\delta_{mp}=  \lambda(\mathcal{T})^2 \ell_d(t) f \Delta p/\mu^\prime$).
  }
  \label{fig:profiles_mp}
\end{figure}

 Similar profiles are displayed on figure \ref{fig:profiles_mp} (at different times) for the marginally pressurized simulation $\mathcal{T}=0.9$. 
    In the marginally pressurized regime, the rupture remains well behind the fluid pressure front. The effect of the elastic stress transfer is minimal (as well observed on the shear-stress profiles). The rupture is primarily driven by the direct pore-pressure effect decreasing the effective normal stress (Fig.~\ref{fig:profiles_mp}d). 
    Even though the overall rupture patch grows much slower, the effect of the in-situ stress gradient appears more pronounced. This can be clearly observed on the re-scaled slip distribution on Fig.~\ref{fig:profiles_mp} - where the deviations to the symmetric solution are more pronounced than for the critically stressed case.

    The larger effect of the in-situ stress gradient is rooted in the fact that in the marginally pressurized regime, the activation of the secondary slipping patch near $x=-L_c$ occurs when the fluid-pressure front reaches $L_c$, i.e. for dimensionless time $\bar{t}\approx 1$ whereas it is triggered by stress transfer (and much earlier) in the critically stressed regime. 
    
\subsection{Development of the relative rupture  asymmetry in both regimes}

    The evolution of the up and down dip amplification factors can be seen in Fig.~\ref{fig:lambdam_lampdap_t} as a function of dimensionless time $\bar{t}$. 
     From figure \ref{fig:lambdam_lampdap_t}, we observe that $\langle\lambda\rangle=(\lambda_{-}+\lambda_{+})/2$ remains close to the symmetric solution for most times (up to an acceleration phase associated with the up-dip tip reaching $L_c$, and the nucleation of the secondary patch). 
      It is interesting to recast the development of the rupture asymmetry by plotting the relative metric $(a_{-}-a_{+})/\langle a\rangle=(\lambda_{-}-\lambda_{+})/\langle\lambda\rangle=\Delta \lambda/\langle\lambda\rangle$ as function of dimensionless time.
 
\begin{figure}
  \centering
  \includegraphics[width=0.8\textwidth]{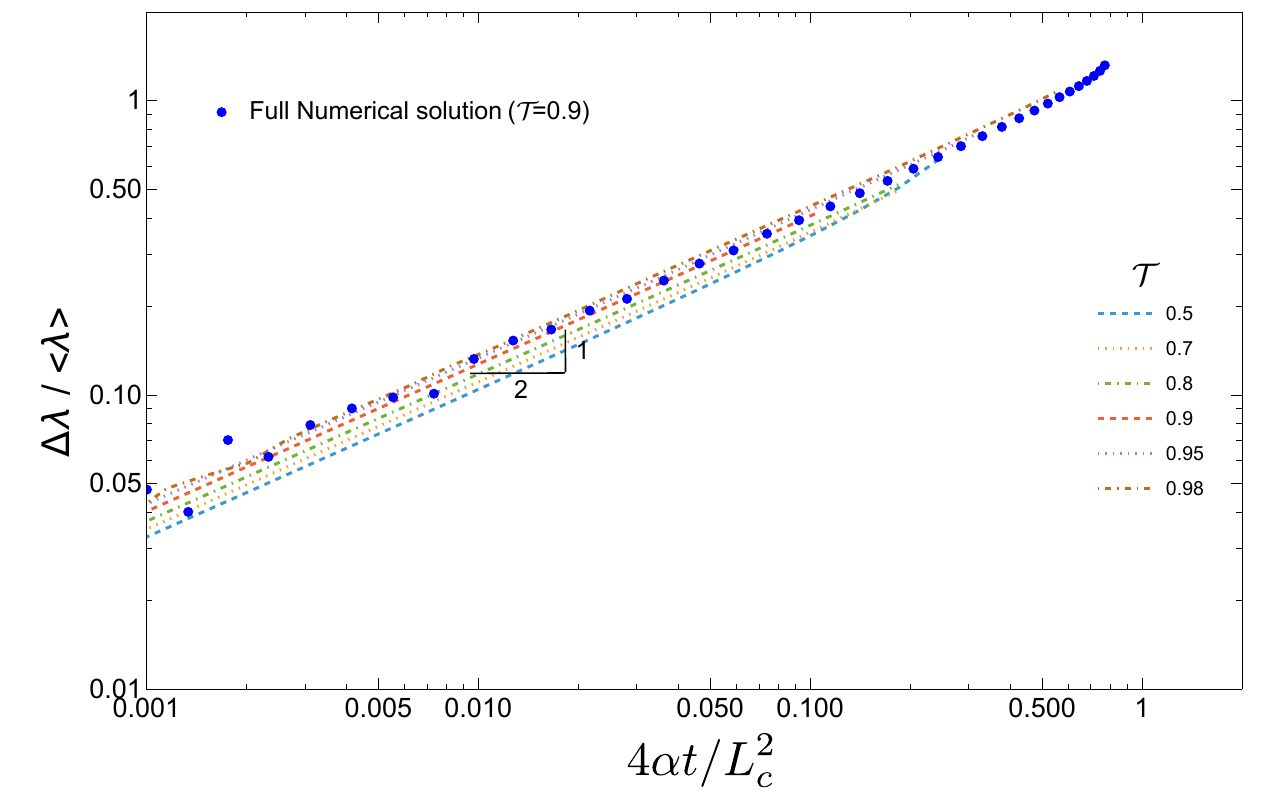}
  \caption{
  Marginally-pressurized regime $\mathcal{T}\gtrsim 0.7$.
  Time evolution of the scaled asymmetry of the amplification factors $\Delta \lambda/\langle\lambda\rangle=2 (\lambda_{-}-\lambda_{+})/(\lambda_{-}+\lambda_{+})$.
 Note the re-scaled time.
  The coupled hydro-mechanical numerical results for $\mathcal{T}=0.9$ are displayed as dots, the corresponding solution of the coupled LEFM propagation conditions are displayed as continuous lines.  }
  \label{fig:asym_mp}
\end{figure}

\paragraph{Marginally pressurized limit}
    Figure \ref{fig:asym_mp} presents the results in the marginally pressurized regime, where the solutions of the LEFM propagation conditions for different values of $\mathcal{T}$ are presented as well as the numerical results for $\mathcal{T}=0.9$.
    We first observe that $\Delta \lambda /\langle\lambda\rangle$ evolves as  $\sqrt{\bar{t}}$ at early times  (for $\bar{t}<0.5$) as anticipated from scaling consideration. An acceleration is visible as the rupture approaches $L_c$ up-dip.
    It is also interesting to note that the curves for $\mathcal{T}\geq 0.9$ are very close to one another. A fit of the numerical results gives $\Delta \lambda/\langle\lambda\rangle\approx 1.1 \sqrt{\bar{t}}$.

    Because $\Delta \lambda \propto \sqrt{\bar{t}}$ (for $\bar{t}<0.5$), the absolute asymmetry of the rupture $(a_{-}-a_{+})$ evolves linearly in time.

 \paragraph{Critically  stressed limit}

\begin{figure}
  \centering
  \includegraphics[width=0.75\textwidth]{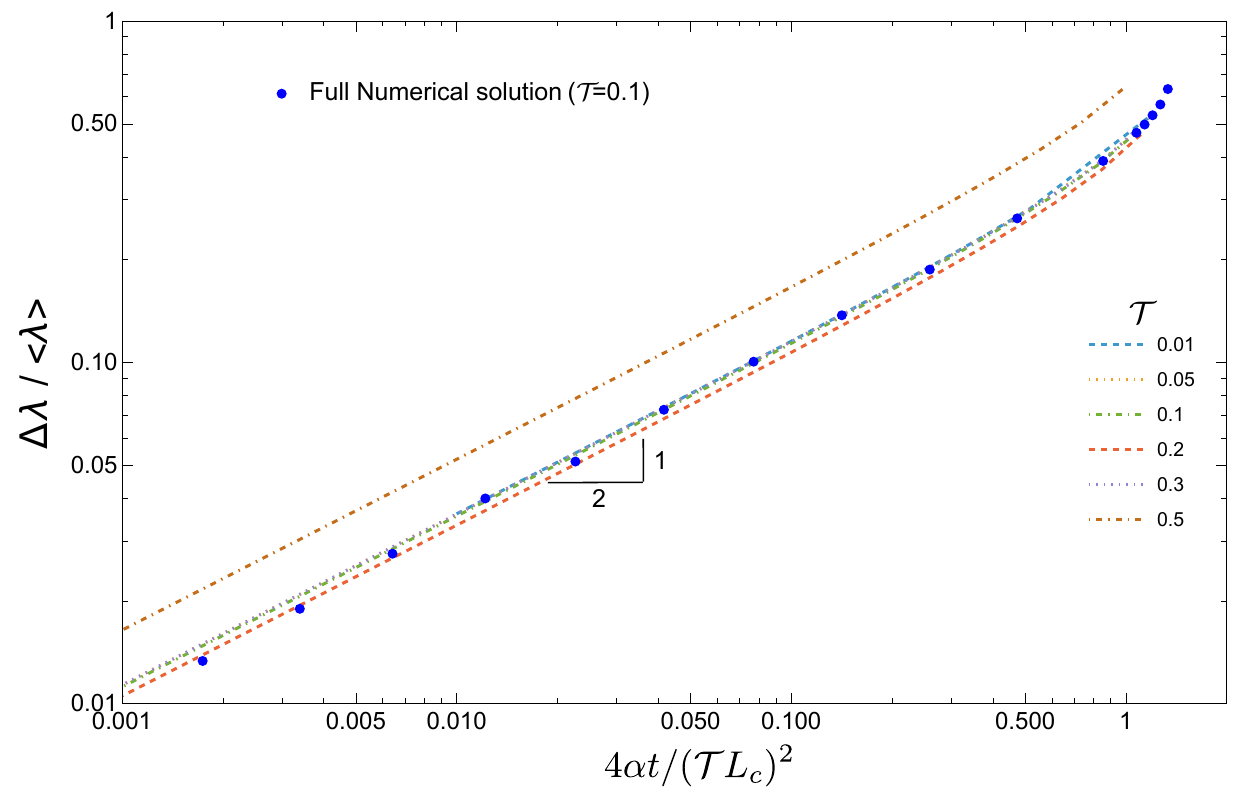}
  \caption{
  Critically-stressed regime $\mathcal{T}\lesssim 0.3$.
  Time evolution of the scaled asymmetry of the amplification factors $\Delta \lambda/\langle\lambda\rangle=2 (\lambda_{-}-\lambda_{+})/(\lambda_{-}+\lambda_{+})$.
 Note the re-scaled time.
  The coupled hydro-mechanical numerical results for $\mathcal{T}=0.1$ are displayed as dots, the corresponding solution of the coupled LEFM propagation conditions are displayed as lines.  }
    \label{fig:asym_cs}
\end{figure}

    Figure \ref{fig:asym_cs}  presents the results in the critically stressed regime, this time rescaling the time  with the characteristic time for the rupture patch half-length to be of order $L_c$: $L_c^2 \mathcal{T}^2/ (4 \alpha) \rightarrow \hat{t}=4\alpha t / (\mathcal{T} L_c)^2$ (see subsection \ref{sec:cs_scale}). 
    Here again, we observe a nice collapse for the evolution of the relative asymmetry $\Delta \lambda / \langle\lambda\rangle$ for values of the fault-stress injection parameter $\mathcal{T}$ below 0.3.
    $\Delta \lambda /\langle\lambda\rangle$ evolves as $\sqrt{\hat{t}}$   at early times  (for $\hat{t}<0.5$) - a fit of the slope provides $\Delta \lambda /\langle\lambda\rangle\approx 0.336 \sqrt{\hat{t}}$ (for $\hat{t}<0.5$).

\subsection{Field-scale illustration: hydraulic stimulation at Rittershoffen}

To place the previous results in a realistic geothermal context and to assess the relevance of stress-gradient effects at kilometer depth, we consider the hydraulic stimulation performed in the Rittershoffen deep geothermal reservoir (France).

\begin{figure}[H]
\centering
\includegraphics[width=0.9\textwidth]{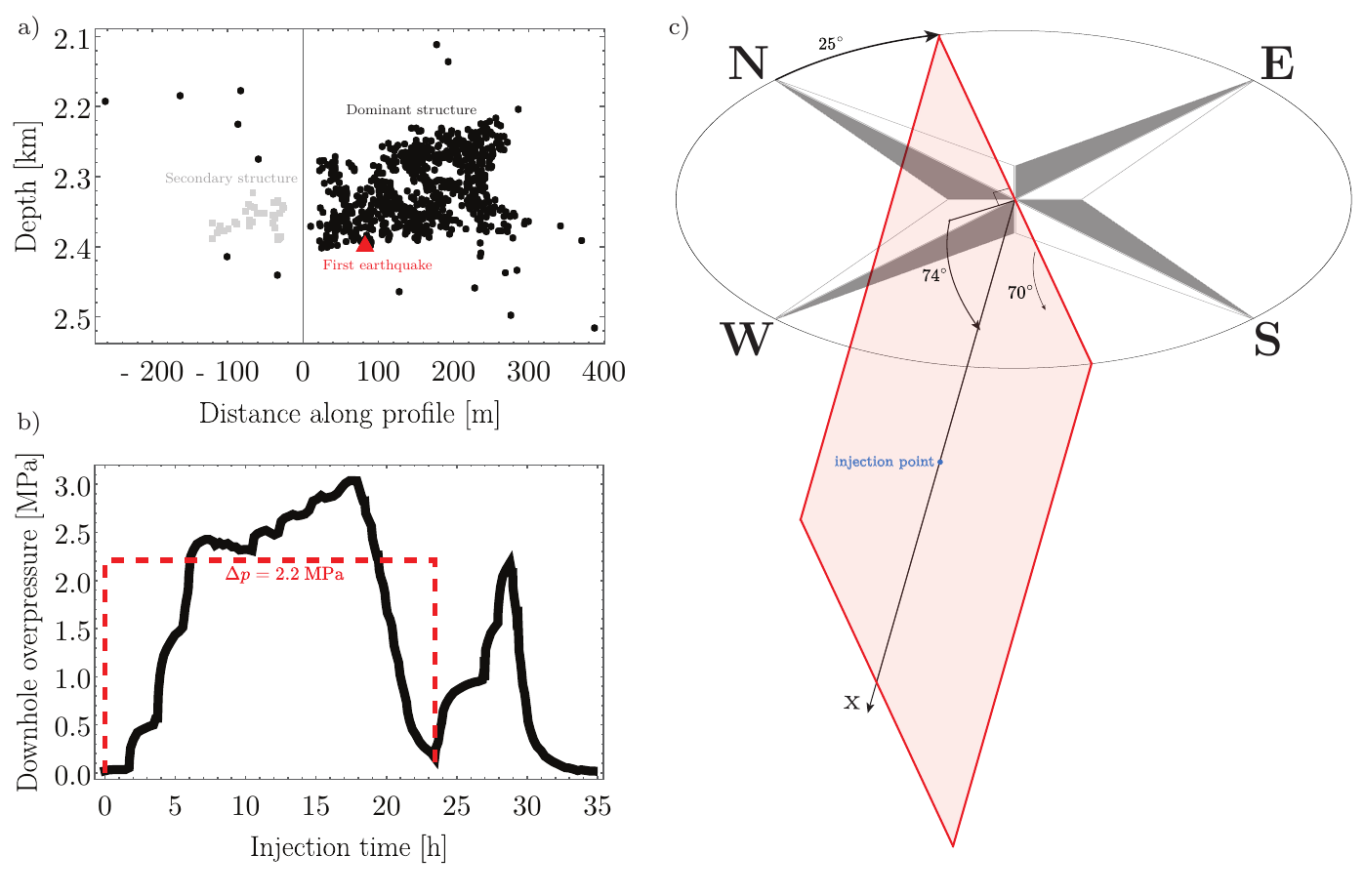}
\caption{(a) Relocated seismic events projected onto the vertical plane with azimuth N025E that struck during the stimulation phase. Figure adapted from \citet{lenglineSeismicityRelatedHydraulic2017a}. (b) Injection downhole overpressure adapted from \citet{baujardHydrothermalCharacterizationWells2017}. The red dashed line corresponds to the root mean square of the pressure values from the beginning of the stimulation to the first pressure drawdown. (c) Semi-analytical model set-up. Best fitting plane with N025E strike azimuth and 74W dip. The model coordinate $x$ is taken along the down-dip direction of this plane, while the resolved shear traction is evaluated along the observed slip direction (rake $\approx 160^\circ$).}
\label{fig:imageD1}
\end{figure}

In June 2013, the ECOGI joint venture performed hydraulic stimulation of the well GRT-1, drilled to a depth of approximately 2580~m with a 640~m open-hole section. Injection lasted for about one day and was accompanied by intense microseismic activity. Although two quasi-parallel structures were activated, the majority of induced events occurred on a single pre-existing fault intersecting the borehole at approximately 2368~m depth and accounting for roughly two-thirds of the injected flow \citep{vidalPrePoststimulationCharacterization2016b}. This structure was reactivated in right-lateral strike-slip motion, with seismicity showing a systematic upward migration during stimulation. Relocated events define a planar structure with strike N025E and dip $74^\circ$W \citep{lenglineSeismicityRelatedHydraulic2017a}. The resulting microseismic cloud extends over roughly 300~m along strike and about 200~m in the vertical direction (Fig.~\ref{fig:imageD1}).

Direct measurements of the in-situ stress profile at Rittershoffen are not available. However, the site lies only about 7~km from the Soultz-sous-Forêts geothermal field, where the stress field has been extensively characterized. Since the distance between the two sites is small and for deep crystalline rocks the crucial role in stress variation with depth is played by long-term formation rheology rather than the geometry of major faults, we assume that principal stress directions and magnitudes are the same for both sites \citep{cornetHowCloseFailure2007b,valleyStressStateSoultz2007,lenglineSeismicityRelatedHydraulic2017a}. Using this model, the principal stresses are taken as
\begin{equation}
\begin{aligned}
S_v &= 33.8 + 0.0255\,(z[\mathrm{m}] - 1377),\\
S_H &\approx S_v,\\
S_h &= 0.54\,S_v,\\
P   &= 0.9 + 0.0098\,z[\mathrm{m}],
\end{aligned}
\end{equation}
where stresses and pore pressure are expressed in MPa. Projecting this stress tensor onto the best-fitting fault plane yields shear tractions of approximately 4.8~MPa along dip and 12~MPa along strike at the injection depth. The resulting rake angle of about $160^\circ$ indicates predominantly right-lateral strike-slip motion on the activated structure \citep{lenglineSeismicityRelatedHydraulic2017a}. To connect the field observations with the stress-gradient scaling, we define the model coordinate $x$ along the down-dip (depth-increasing) direction on the best-fitting fault plane. The resolved shear traction $\tau$ is evaluated along the observed slip direction (rake), while invariance is assumed in the out-of-plane direction as a two-dimensional reduction. Note that the slip is then mostly perpendicular to $x$, i.e. closer to anti-plane than to in-plane shear. This does not affect the predicted tip positions: the propagation conditions \eqref{ch3:7}--\eqref{ch3:8} do not involve the elastic modulus $\mu'$, which only enters the slip distribution. With this convention, the gradients of resolved tractions along the fault are estimated as
\begin{equation}
m_{\sigma'} = 7.5\times10^{3}\ \mathrm{Pa/m},
\qquad
m_{\tau} = 5.1\times10^{3}\ \mathrm{Pa/m}.
\label{eq:grad_ritt}
\end{equation}

To estimate the stress-gradient length scale $L_c$, we first estimate the fault-stress injection parameter at the injection point
\[
\mathcal{T}=\frac{f\sigma'_0-\tau_0}{f\Delta p}
\]
to characterize the regime of the injection.
The numerator $f\sigma'_0-\tau_0$ is a local quantity and must be evaluated with care. In the critically stressed limit, which we believe applies at Rittershoffen \citep{cornetHowCloseFailure2007b,lenglineSeismicityRelatedHydraulic2017a}, an uncertainty of order $0.1$~MPa in $f\sigma'_0-\tau_0$ produces large variations in predicted rupture-tip positions. Moreover, the regional stress model of \citet{cornetHowCloseFailure2007b} describes large-scale trends and must be used with care to infer the local distance to failure at the nucleation point. We therefore constrain $\mathcal{T}$ pragmatically by matching the order of magnitude of the observed microseismic extent with the uniform-stress analytical solution of \citet{viescaSelfsimilarFaultSlip2021}. For a large (but plausible) fault diffusivity of $\alpha=0.05~\mathrm{m}^2/\mathrm{s}$, a reasonable match is obtained with an amplification factor $\lambda\simeq 1.8$ and $\mathcal{T}\simeq 0.2$. We adopt $f=0.85$ following \citet{lenglineSeismicityRelatedHydraulic2017a} and estimate a characteristic overpressure $\Delta p\simeq 2.2$~MPa from the downhole overpressure record during the stimulation phase. This injection is therefore likely in the critically-stressed regime. 
With these values, the critically stressed stress scale is
\[
\tau_{el} = f\sigma'_0-\tau_0 = \mathcal{T} f\Delta p
          \approx 0.37~\mathrm{MPa}.
\]
From the resolved traction gradients \eqref{eq:grad_ritt} along the best-fitting rupture plane,
the effective strength gradient is
\[
fm_{\sigma'}-m_{\tau}
 \simeq 1.3\times10^{3}~\mathrm{Pa/m}.
\]
The stress-gradient length scale then follows as
\[
L_c=\frac{ f\sigma'_0-\tau_0}{fm_{\sigma'}-m_{\tau}}
   \approx 300~\mathrm{m}.
\]

This value is of the same order as the observed dimensions of the
microseismic cloud at Rittershoffen, which extends over roughly
300~m along strike and about 200~m in the vertical direction.
The proximity of these length scales indicates that rupture growth
during this hydraulic stimulation reached distances for which stress-gradient effects
are expected to start to be noticeable, i.e.\ $a\sim L_c$, but not
to dominate the dynamics of growth. The model predicts a
moderate rupture asymmetry due to stress gradient effects; for the critically stressed case,
the corresponding asymmetry factor is $\Delta a /\langle a\rangle \sim 0.3$. 

\section{Conclusions}\label{sec:conclusions}

Linear variations of the in-situ stresses with depth are ubiquitous, yet models of fluid-driven aseismic slip have so far assumed uniform tractions along the fault. We have extended the plane-strain model of fluid-driven frictional rupture under constant overpressure to linear gradients of the in-situ shear and effective normal tractions. The condition of vanishing stress intensity factors at both rupture tips provides two coupled equations for the up-dip and down-dip tip positions. We have solved them semi-analytically and compared their solution with coupled hydro-mechanical simulations performed with the solver of \citet{LeBr27}. Both approaches agree up to the activation of a secondary slipping patch. This secondary patch nucleates at $x=-L_c$, where the in-situ distance to failure of the fault vanishes: the fault there is initially at yield, so that any additional loading triggers slip.

The stress gradient notably breaks the self-similarity of the uniform-stress solution. Rupture growth is governed by the fault-stress injection parameter $\mathcal{T}$ at the injection point and by the ratio of the diffusion length $\ell_d=\sqrt{4\alpha t}$ to the stress-gradient length scale $L_c$. The latter is the ratio between the initial distance to failure at the injection point and its gradient along the fault. Growth is symmetric at early times and follows the uniform-stress solution of \citet{viescaSelfsimilarFaultSlip2021}. Asymmetry then develops in both regimes: the up-dip tip runs ahead of the symmetric solution, whereas the down-dip tip lags behind it.

In the marginally pressurized regime ($\mathcal{T}\rightarrow 1$), rupture is controlled by the direct pore-pressure effect and lags behind the pressure front. The relevant dimensionless time is $\bar t=4\alpha t/L_c^2$, and the relative asymmetry grows as $\Delta\lambda/\langle\lambda\rangle\approx 1.1\sqrt{\bar t}$. As the pressure front approaches $L_c$, the up-dip tip accelerates markedly and the slip profile becomes strongly skewed.

In the critically stressed regime ($\mathcal{T}\ll 1$), rupture is driven by elastic stress transfer and outpaces diffusion by a factor $\lambda\propto 1/\mathcal{T}$. The relevant time is then $\hat t=\bar t/\mathcal{T}^2$, and the relative asymmetry grows as $\Delta\lambda/\langle\lambda\rangle\approx 0.34\sqrt{\hat t}$. The up-dip tip reaches $L_c$ earlier than in the marginally pressurized regime. The asymmetry therefore remains moderate: the ratio of up-dip to down-dip rupture lengths stays below two, and the slip profile remains nearly symmetric.

In both regimes, the single-patch solution thus ceases to be valid once the loading reaches $x=-L_c$. The secondary patch is triggered there by elastic stress transfer in the critically stressed regime, and by pore-pressure diffusion in the marginally pressurized one.

For practical field conditions, $L_c$ is readily estimated from the profile of the initial distance to failure along the fault, $L_c=(f\sigma'_0-\tau_0)/\partial_x(f\sigma'_0-\tau_0)$. Comparing it with the expected rupture extent directly gauges the importance of the in-situ stress gradient for a given injection. For the 2013 stimulation at Rittershoffen, $L_c$ is of the same order as the microseismic cloud, and the model predicts a moderate asymmetry, $\Delta a/\langle a\rangle\sim 0.3$.

The analysis presented here has a number of  simplifying assumptions. First, we restricted our discussion to the evolution of the primary slipping patch. The multi-patch evolution beyond that time is a topic worth further investigation. 
Second, the case of sudden variations of in-situ stress associated with lithological changes is likely more relevant in practice. The stress-gradient scale $L_c$ (either for the critically stressed or marginally pressurized regime) should be compared with the extent of lithological units to gauge its importance. 
Third, accounting for frictional weakening as well as dilatancy in the presence of a stress gradient still needs to be properly investigated. Notably the effect of dilatancy and permeability variation is likely as - if not more - important in practice than initial stress variation. Finally, the extension to three-dimensional ruptures under constant injection rate is worth pursuing, where the rupture elongation in the rake direction associated with Poisson's ratio may compete with or reinforce the effect of the in-situ stress gradient.
\paragraph{Declaration of generative AI and AI-assisted technologies in the writing process\\}
During the preparation of this work the authors used Opus 5.5 to improve language and readability of part of the text. After using this tool/service, the authors reviewed and edited the content as needed and take full responsibility for the content of the publication.

\paragraph{CRediT author statement\\}   
\textbf{R. Fakhretdinova}: Conceptualization, Validation, Formal analysis, Visualization, Writing - Original Draft.
\textbf{B. Lecampion}: Conceptualization, Methodology, Software, Validation, Formal analysis, Writing - Original Draft, Visualization, Resources, Supervision, Funding acquisition.

\paragraph{Acknowledgments}
 R.F. was partially funded by the Federal Commission for Scholarships for Foreign Students via a Swiss Government Excellence Scholarship.

\bibliographystyle{plainnat}
\bibliography{references}

\end{document}